\documentclass[superscriptaddress,
preprint,
amsmath,amssymb,
aps,
]{revtex4-1}

\usepackage{graphicx}
\usepackage{dcolumn}
\usepackage{float}
\usepackage{subfigure}
\usepackage{color}
\usepackage{url}

\begin{document}

\title{Equilibrium reconstruction with 3D eddy currents in the Lithium Tokamak eXperiment}

\author{C.~Hansen}
\email{hansec@uw.edu}
\affiliation{Department of Aeronautics \& Astronautics, University of Washington, Seattle, Washington 98195, USA}
\affiliation{Department of Applied Physics \& Applied Math, Columbia University, New York, New York 10027, USA}
\author{D.~P.~Boyle}
\affiliation{Princeton Plasma Physics Laboratory, Princeton, New Jersey 08543, USA}
\author{J.~C.~Schmitt}
\affiliation{Physics Department, Auburn University, Auburn, Alabama 36849, USA}
\author{R.~Majeski}
\affiliation{Princeton Plasma Physics Laboratory, Princeton, New Jersey 08543, USA}

\date{\today}

\makeatletter{}\begin{abstract}
Axisymmetric free-boundary equilibrium reconstructions of tokamak plasmas in the Lithium Tokamak eXperiment (LTX) are performed
using the PSI-Tri equilibrium code. Reconstructions in LTX are complicated by the presence of long-lived non-axisymmetric eddy
currents generated by vacuum vessel and first wall structures. To account for this effect, reconstructions are performed with
additional toroidal current sources in these conducting regions. The source distributions are fixed poloidally, but their scale
is adjusted as part of the full reconstruction. Eddy distributions are computed by toroidally averaging currents, generated by
coupling to vacuum field coils, from a simplified 3D filament model of important conducting structures. The full 3D eddy
current fields are also used to enable the inclusion of local magnetic field measurements, which have strong 3D eddy current
pick-up, as reconstruction constraints. Using this method, equilibrium reconstruction yields good agreement with all
available diagnostic signals. An accompanying field perturbation produced by 3D eddy currents on the plasma surface with
primarily $n=2$, $m=1$ character is also predicted for these equilibria.
\end{abstract}

\maketitle

\section{Introduction}
\makeatletter{}Reconstruction of plasma equilibria is crucial to the understanding of transient dynamics, transport and performance in
magnetic confinement experiments. In nominally axisymmetric devices, such as the tokamak, this typically involves the
assumption of an axisymmetric, ideal magnetohydrodynamic (MHD) equilibrium, yielding the Grad-Shafranov equation\cite{Grad1958}. Tools to solve the
inverse problem, to find the equilibrium whose reconstructed diagnostic measurements best match the experimentally
observed values, are routinely used by many experiments\cite{Lao1990,Sabbagh2001,Brix2008}. However, reconstructions
can be complicated by the presence of eddy currents induced in conducting device structures, such as the first
wall\cite{Berzak2010,Berzak2012}, vacuum vessel\cite{Anderson2004} and coil supports\cite{Ma2015}, which are often
poorly diagnosed in contrast to the plasma itself.

A new method, presented here, was applied to capture the effect of both axisymmetric and 3D eddy currents on axisymmetric plasma
equilibria in the Lithium Tokamak eXperiment (LTX). LTX is an Ohmically-heated spherical tokamak (nominal parameters: $R_0$~=~0.4~m,
$a$~=~0.26~m, $\kappa$~=~1.6, $B_0$~=~0.17~T) with a close-fitting low-recycling wall composed of thin lithium
coatings evaporated onto a segmented metallic shell\cite{Majeski2009}. These shell segments support long-lived eddy currents with
decay time scales comparable to the plasma lifetime ($\sim$~25~ms). The vacuum vessel (VV) has higher resistivity but
contains toroidally continuous current paths, supporting similarly long-lived eddy currents. Together, eddy currents in the
shell and VV strongly modify the vacuum field in the plasma region and must be taken into account during reconstructions.
Additionally, toroidal and poloidal breaks in the shell
force 3D current paths, which in turn produce 3D field pickup on local magnetic field probes that must be accounted
for to allow their use as reconstruction constraints. This effect is enhanced due to the required placement of magnetic
probes at the center of the shell breaks where the 3D effects are largest\cite{Schmitt2014}.

By supplementing a traditional Grad-Shafranov based reconstruction with toroidal current sources of specified shape but
variable amplitude in the shell and VV regions, good agreement was achieved between observed and reconstructed
signals in LTX. Toroidal current sources for eddy regions were determined using a filament model for each conducting
region: toroidal loops for the vacuum vessel and a 3D mesh of loops for the shell. The best results were obtained
using four total eddy current sources: two current sources for the VV, corresponding to the longest-lived vertically symmetric
eigenmodes from an inductive-resistive (L-R) model, and a single current source for each of the upper and lower shell segments, corresponding to current induced
by coupling to the Ohmic-heating (OH) coil in the resistive limit. For the shell, the toroidally averaged current was used as a current source and the full 3D
field was used to correct the field for comparison with the local magnetic probes. This model assumes that 3D effects on the
plasma are small so that an axisymmetric model is still a valid approximation for the plasma equilibrium. A significant reduction
in the reconstruction error was observed with the addition of this eddy current model compared to both reconstruction without
an eddy current model and reconstruction with an eddy current model but without 3D correction of the local field probes.
This paper will describe this new reconstruction method and provide results for application to equilibria of interest in LTX.

The remainder of this paper will be organized as follows: In section \ref{sec:psi_tri} the PSI-Tri code, which was used for
this investigation, will be briefly described. In section \ref{sec:model} the model used to approximate vacuum vessel and shell
eddy currents will be described. Section \ref{sec:results} will present application of this method to reconstruction of LTX
equilibria. Finally, the results of this work will be summarized and the direction of future work will be discussed
in section \ref{sec:discussion}.

\section{PSI-Tri} \label{sec:psi_tri}
\begin{figure}[]
  \begin{center}
   \includegraphics[width=.3\linewidth]{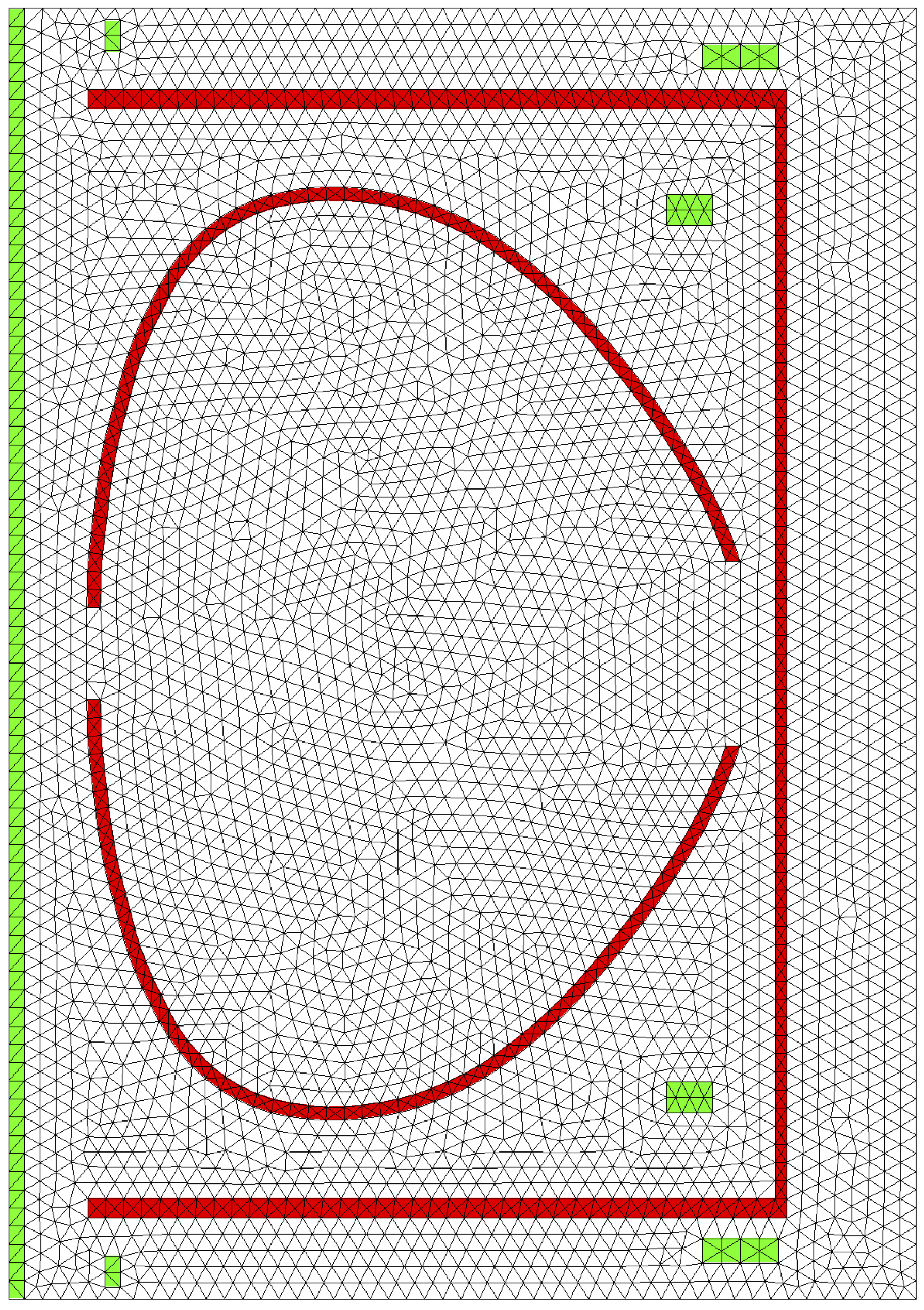}
  \end{center}
\caption{Mesh used for LTX equilibrium reconstructions showing in-mesh vacuum field coils (green)and passive conducting (red) regions.
         Additional vacuum field coils exist outside the mesh that are included through boundary conditions.}
\label{fig:ltx_mesh}
\end{figure}

\makeatletter{}The PSI-center Triangular mesh code (PSI-Tri) is a 2D high order, finite element, free-boundary equilibrium and reconstruction
code. An interface to the CUBIT\cite{cubit} and T3D\cite{t3d} meshing packages allows generation of computational grids
directly from computer-aided design (CAD) models. The mesh used for LTX reconstructions, which was generated using CUBIT and
is illustrative of these features, is shown in figure \ref{fig:ltx_mesh}. The Grad-Shafranov equation is discretized using
a Galerkin finite element method with uniformly spaced nodal basis sets (Lagrange); second order elements were used for
the following results. Parallelization, using OpenMP\cite{openmp}, is used throughout the code to efficiently utilize
modern multicore processors.

The Grad-Shafranov equation in PSI-Tri is solved using an under-relaxed Picard iteration, with linear solves on each
iteration computed using either direct (LU) or multigrid preconditioned iterative solvers (Conjugate-Gradient). Vacuum field
coils can be defined externally, via boundary conditions, or internally, through meshed regions with defined current profiles.
Free-boundary equilibria are converged using alternating fixed boundary and free boundary iterations. Boundary poloidal flux
due to internal currents can be computed using either a fast boundary solution method\cite{JardinBook} or a more accurate, but
slower, direct integral method. Active position control is supported using total plasma pressure for radial position and
feedback controlled in-mesh coils for vertical position. Both limited and diverted topologies are supported.

Equilibrium reconstruction in PSI-Tri is performed using the Levenburg-Marquadt method to minimize the error between
reconstructed and observed diagnostic measurements defined by $\chi^2$ (eq. \ref{eq:chi_sq}). This method was chosen
to maintain flexibility in the formulation of flux functions and constraining diagnostics, in contrast to reconstruction
techniques that employ linear least-squares fitting on each step of the equilibrium Picard iteration\cite{Lao1990}. In
particular, this flexibility allows the use of inherently non-linear relationships between constraints and parameters,
such as the equilibrium-defined flux functions used for modeling driven equilibria in HIT-SI\cite{Jarboe2012,HansenThesis}.

\begin{equation} \label{eq:chi_sq}
\chi^2_i = \frac{(f^{obs}_i - f^{recon}_i)^2}{\sigma_i^2}
\end{equation}

\section{Eddy-Current Model} \label{sec:model}
\makeatletter{}The eddy currents induced in LTX were studied previously with time-domain simulations using
3D thin wall models\cite{Berzak2010,Schmitt2014}. These studies have shown good agreement with data
on vacuum shots, when the eddy currents are driven by well known sources (fixed coils). However,
reproducing the correct eddy current distribution during a plasma shot requires accurate knowledge
of the distribution of current in the plasma, i.e. an equilibrium, which itself requires the effect
of eddy currents. As a result, in order to perform accurate reconstructions two approaches have
been considered: 1) A coupled method, where an eddy current simulation of the discharge is
performed and reconstructions are done at fixed time intervals, to update the current distribution
in the plasma, taking the eddy currents as known. 2) A decoupled method where eddy current
simulations are used to determine a reduced set of eddy currents that can be included directly as
free-parameters in the reconstruction. The second method, which is presented here, has several
advantages. Primarily, it can be performed at a single time point without processing
the entire discharge -- a useful trait for inter-shot reconstructions to help guide experimental
operation. For discussion of further work on both methods, see section \ref{sec:discussion}.

\begin{figure}[]
  \begin{center}
   \includegraphics[width=.3\linewidth]{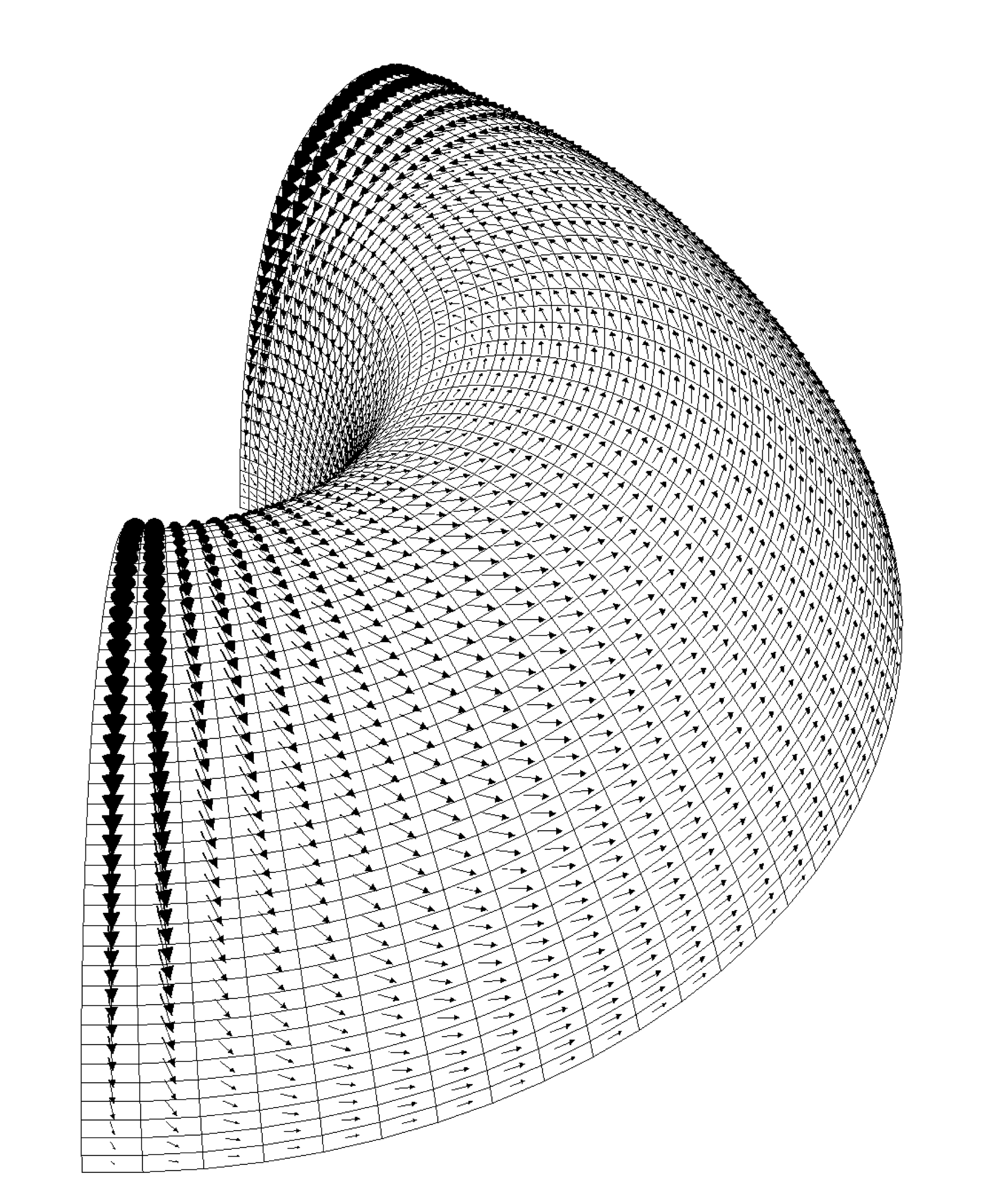}
  \end{center}
\caption{3D eddy current distribution in a single shell segment used in reconstructions.}
\label{fig:ltx_eddy}
\end{figure}

Thin-wall models have proven successful for studying eddy currents in LTX and other experiments\cite{Schmitt2014,Bialek2001,Okabayashi2005,Sabbagh2010}.
However, extracting information from numerical tools that perform time-dependent simulations for integration
into an axisymmetric model is difficult, depending on the spatial discretization and other factors. In
particular, when used in a Grad-Shafranov (G-S) equilibrium, the toroidally averaged eddy currents in a given
region must be known. As a result, a further simplification to a meshed filament model of the LTX shell
was used in this investigation. To produce this model in a way that is compatible with the G-S solver, the LTX wall
regions (VV and shell, shown as the red regions in figure \ref{fig:ltx_mesh}) were first meshed using quadrilaterals (quads).
Node points were then placed at the center of each quad, which were in turn used to define a filament mesh for eddy
currents and used to subdivide the quads into 4 triangles producing a uniformly triangular grid for PSI-Tri. For
axisymmetric regions, like the VV, each node point was used to define the location of a toroidal filament,
which together form a filament model for that region. For regions that are not toroidally contiguous, like the shell,
each node point was used to define the poloidal node locations of a grid of quads generated by extruding
the poloidal nodes in toroidal steps out to the toroidal extent of the conducting region, shown in figure
\ref{fig:ltx_eddy}. Eddy currents are then represented by allowing current to flow as individual loop
currents around each cell, where the conducting area of each cell is lumped onto its edges. With a
filament model in place for both types of conducting regions the eddy current distributions were then analyzed.

Based on previous analysis of vacuum and plasma shots using VALEN\cite{Bialek2001} it is known
that the VV eddy currents are primarily located at the inboard and outboard corners of the upper and
lower VV walls\cite{Schmitt2014}. This is due to a combination of ports located in the central regions of the
upper, lower and outboard VV walls, which impede toroidal current, and thicker material joints in the corners
which provide a low resistance path. As a result, two VV eddy current distributions were used for
equilibrium reconstruction that correspond to the longest-lived vertically symmetric eigenmodes from an L-R model
based on the filaments in these two highly conducting regions only. Other variations were also investigated,
including a resistivity profile to capture the current distribution\cite{Berzak2010} and external coil
coupling as used for the first wall shell. However, the best results were found with the eigenmode
decay model used here.

\begin{figure}[]
  \begin{center}
   \includegraphics[width=.4\linewidth]{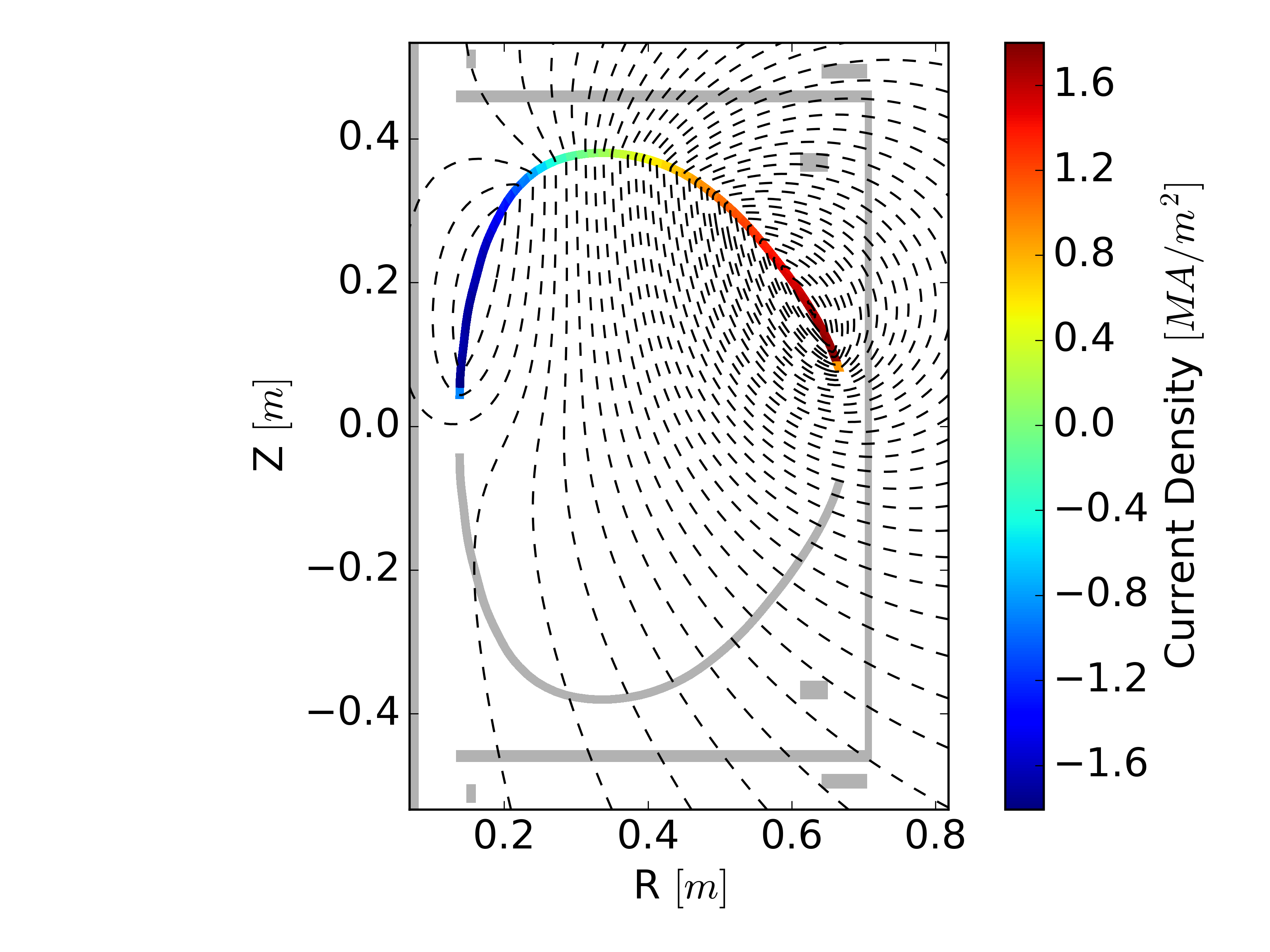}
  \end{center}
\caption{Toroidal current source and corresponding poloidal flux contours for the upper shell segments created by
         toroidally averaging figure \ref{fig:ltx_eddy}. The current source is shown with the same
         amplitude (but with a different color scale) as in the equilibrium given in figure \ref{fig:ltx_psi},
         which corresponds to approximately $6~kA$ of circulating current in each of the upper shell segments.}
\label{fig:ltx_vac_psi}
\end{figure}

The shell is more complicated to treat due to its 3D nature. As with the VV, several different
current distribution types and quantities were tried. All of these were based on an L-R model of the
3D loop filament mesh shown in figure \ref{fig:ltx_eddy}. For this paper, results are presented using
a single mode for each shell segment, produced in the resistive limit (equation \ref{eq:res_limit}, where $\textbf{V}_{OH}$ is
the loop voltage produced around each filament by the Ohmic transformer). The resistivity matrix $R$ defines
the voltage drop around each loop due to its current and current in the adjacent loops in the mesh.
\begin{equation} \label{eq:res_limit}
\textbf{I} = R^{-1} \textbf{V}_{OH}
\end{equation}
The corresponding current distribution is shown in figure \ref{fig:ltx_eddy}. The toroidally averaged
toroidal current source and resulting poloidal flux for this mode used in equilibrium reconstruction are
shown in figure \ref{fig:ltx_vac_psi}. Due to the construction of the grid, generated by revolving a
set of poloidal nodes, averaging the toroidal current to each node point is straightforward. As the
shell has 3D currents, in addition to the axisymmetric source terms, local field corrections are also
needed. These are computed by evaluating the fields at each probe location due to the full 3D eddy
current fields $\textbf{B}^{eddy,3D}_i$ and the fields predicted by solving the G-S equation with only the
toroidally averaged eddy source term $\textbf{B}^{eddy,2D}_i$, as in figure \ref{fig:ltx_vac_psi}. The full
reconstructed field at the probe $\textbf{B}^{recon,3D}$ is then given by equation \ref{eq:eddy_corr},
\begin{equation} \label{eq:eddy_corr}
\textbf{B}^{recon,3D} = \textbf{B}^{recon,2D} + W_i \left( \textbf{B}^{eddy,3D}_i - \textbf{B}^{eddy,2D}_i \right)
\end{equation}
where $\textbf{B}^{recon,2D}$ is the field from the G-S reconstruction and $W_i$ is the amplitude of the corresponding
eddy current source $i$ in the reconstruction.

\section{Results} \label{sec:results}
\makeatletter{}The reconstruction method presented above was applied to a discharge in LTX where
Thomson scattering was used to measure the electron pressure profile\cite{BoyleThesis,Majeski2017,Boyle2017}. Of particular
interest was the
plasma state late in time ($\sim$~474~ms) when flat temperature profiles were observed\cite{Majeski2017,Boyle2017}. A series of
free-boundary reconstructions were performed at this time using PSI-Tri with and without eddy current
models. Plasma toroidal current was parameterized using quadratic polynomials for both the
pressure gradient and toroidal flux source terms, with a non-zero pressure gradient allowed at the
plasma boundary, yielding five free reconstruction parameters for the plasma. At this time in the
discharge a small vertical asymmetry was present in the magnetic signals so a small vertical offset in the
magnetic axis (1.5 cm below the midplane) was used for reconstruction.

\begin{figure}[]
  \begin{center}
   \includegraphics[width=.47\linewidth]{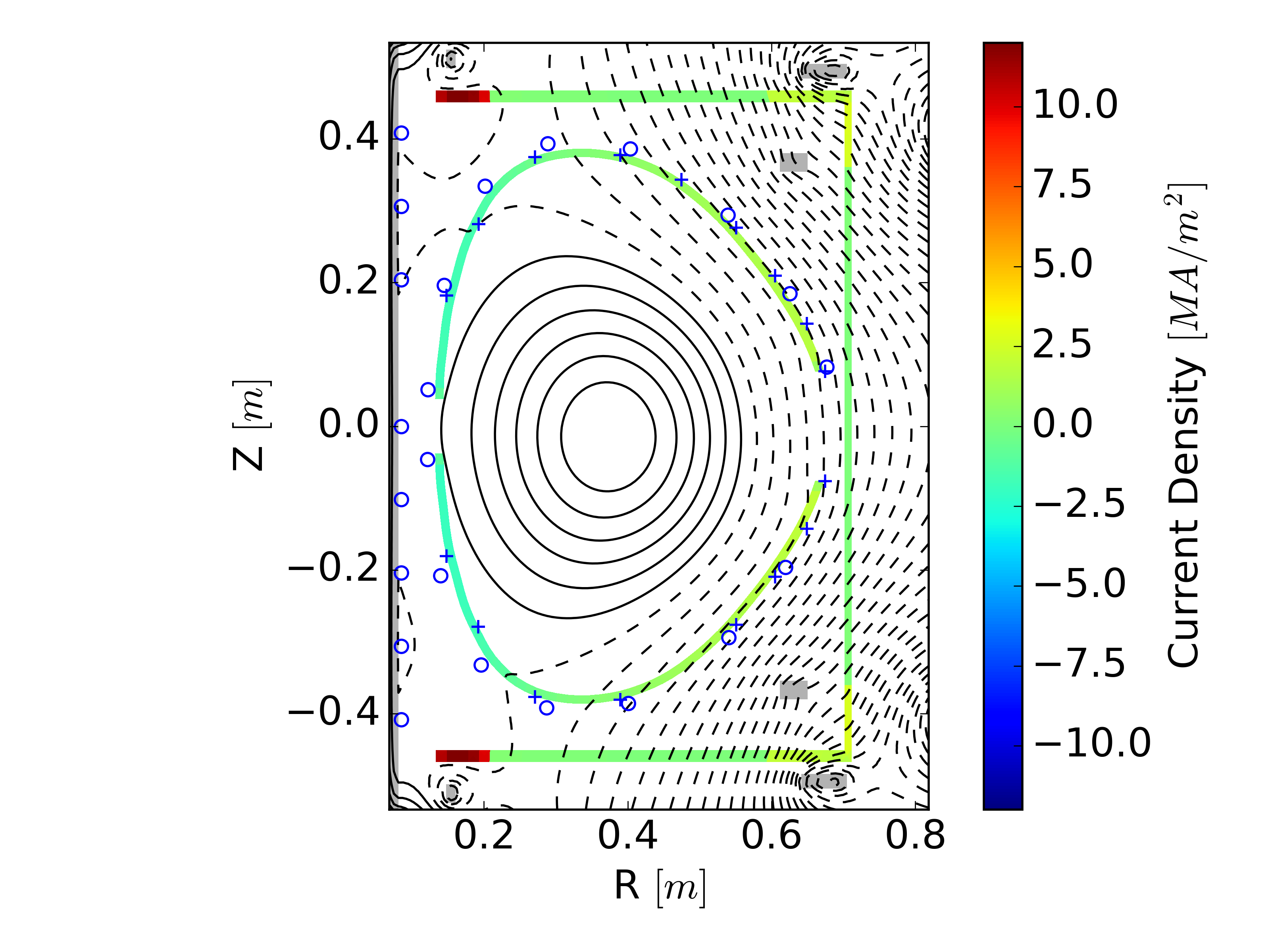}
  \end{center}
\caption{Poloidal flux contours and toroidal eddy currents (shaded regions) from equilibrium reconstruction of
         LTX discharge 1504291321 at 474 ms. The location of magnetic diagnostics used in the reconstruction,
         flux loops (blue circles) and local field probes (blue crosses), are also shown.}
\label{fig:ltx_psi}
\end{figure}

Five sets of diagnostics were used to constrain the reconstructions: 1) Toroidal current as measured
by an internal Rogowski. 2) The diamagnetic flux as measured by a toroidal flux loop. 3) Plasma pressure
as measured at 7 spatial points via Thomson Scattering\cite{Strickler2008,BoyleThesis}. 4) A set of 17 poloidal flux loops located on
the inboard VV wall and around the backside of the upper and lower shell segments. 5) A set of 34 local magnetic field probes
(17 radial and 17 vertical) located at a single toroidal location, in the center of a toroidal shell break, and poloidally distributed
roughly uniformly and in line with the poloidal cross-section of the shell. The poloidal location of magnetic diagnostics are shown
in figure \ref{fig:ltx_psi}; for a detailed discussion of these diagnostics see references \cite{Berzak2010,Berzak2012,Schmitt2014}.
A constraint was also placed on the minimum safety safety factor to exclude the $q=1$ surface from the plasma, resulting in a
total of 61 constraints.

The resulting reconstruction error ($\chi^2$) for each diagnostic set is shown in table \ref{tab:ltx_chi_comp}
for 3 different reconstructions: 1) A standard G-S reconstruction without eddy current effects (No Eddy),
resulting in 5 free-parameters. 2) A fully axisymmetric reconstruction with two eddy current modes in the VV
using the model described in the previous section (VV Only), resulting in 7 free-parameters. 3) A
reconstruction using the full model with an axisymmetric plasma, two eddy current modes in the VV, and one
eddy current mode in each of the upper and lower shell segments, resulting in 9 free-parameters. The reconstruction
error was reduced on all signal sets with the addition of more sophisticated eddy current models, with the exception
of a slight increase in the poloidal flux error when shell currents were added. The very high error shown in table
\ref{tab:ltx_chi_comp} for the No Eddy case highlights the importance of eddy currents in reconstructing plasma
equilibria in LTX.

\begin{table}
 \caption{$\chi^2$ for each diagnostic set for 3 different reconstructions: 1) baseline case with no eddy currents,
 2) case with vacuum vessel eddy currents only, 3) case with vacuum vessel and shell eddy currents. The number of
 free-parameters for each reconstruction and the number of diagnostic signals in each set are also provided.}
 \label{tab:ltx_chi_comp}
 \begin{center}
 \begin{tabular}{ | l | c | c | c | }
   \hline
   Diagnostic Set & No Eddy (5) & VV Only (7) & VV \& Shell (9) \\ \hline
   Toroidal Current (1)   & 256.2  & 0.6   & 0.0     \\
   Diamagnetic Flux (1)   & 163.7  & 3.6   & 2.2     \\
   Thomson (7)            & 66.7   & 8.6   & 1.0     \\
   Flux Loops (17)        & 117.0  & 18.8  & 22.1    \\
   Re-Entrant Probes (34) & 1258.2 & 676.3 & 30.0   \\ \hline
 \end{tabular}
 \end{center}
\end{table}

\begin{figure}[]
  \begin{center}
   \includegraphics[width=.47\linewidth]{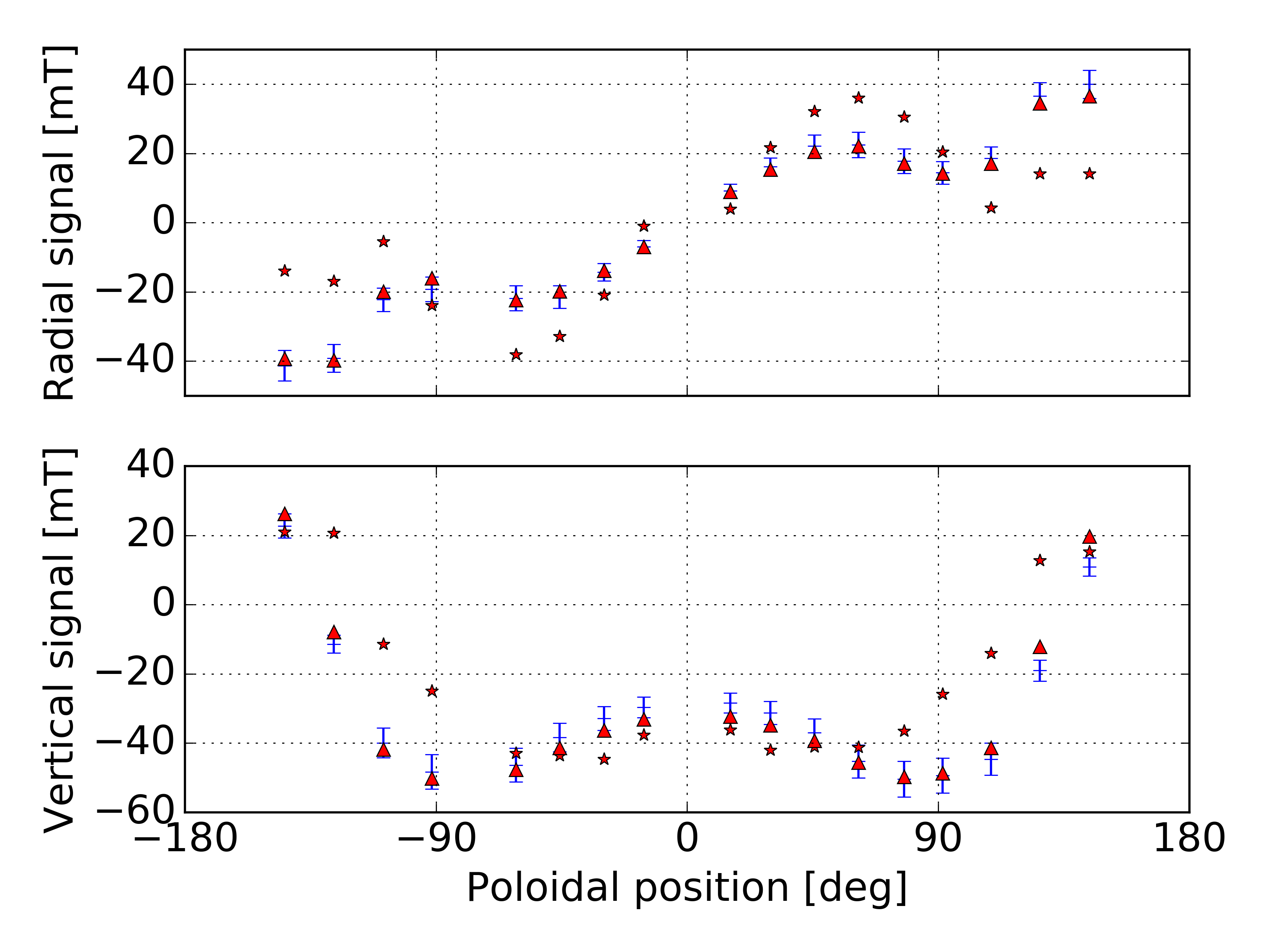}
  \end{center}
\caption{Comparison of reconstructed (red triangles) and observed (blue crosses) local magnetic probe signals in
         LTX at 474 ms in shot 1504291321. Reconstructed signals are also shown without the 3D eddy field
         correction (red stars) to illustrate the magnitude and shape of 3D field contributions to local probe measurements.
         Positive poloidal positions are above the midplane with zero corresponding to the outboard midplane.}
\label{fig:ltx_recon_comp}
\end{figure}

This technique improves upon prior models used on LTX by achieving agreement with local fields measurements
made in the toroidal shell break. Figure \ref{fig:ltx_recon_comp} shows the importance of the 3D field corrections
in reproducing these observed local field signals. Using only toroidally averaged fields (red stars), the
agreement with the observed signals (blue crosses) is poor, while after the eddy current correction is applied
(red triangles) good agreement is achieved. The difference between the pre and post correction fields can
be quite large where the breaks in the shell cause the poloidally flowing current at the ends of the plate
to reinforce each other, as seen in the radial field near the inboard midplane. This correction has a specific
shape and amplitude, which is produced by the recirculating current in the shell shown in
figure \ref{fig:ltx_eddy}. The resulting agreement between observed signals and the full reconstructed field
indicates good agreement between the chosen eddy current distribution and the distribution present in the
experiment at this time.

\begin{figure}[]
  \begin{center}
   \includegraphics[width=.47\linewidth]{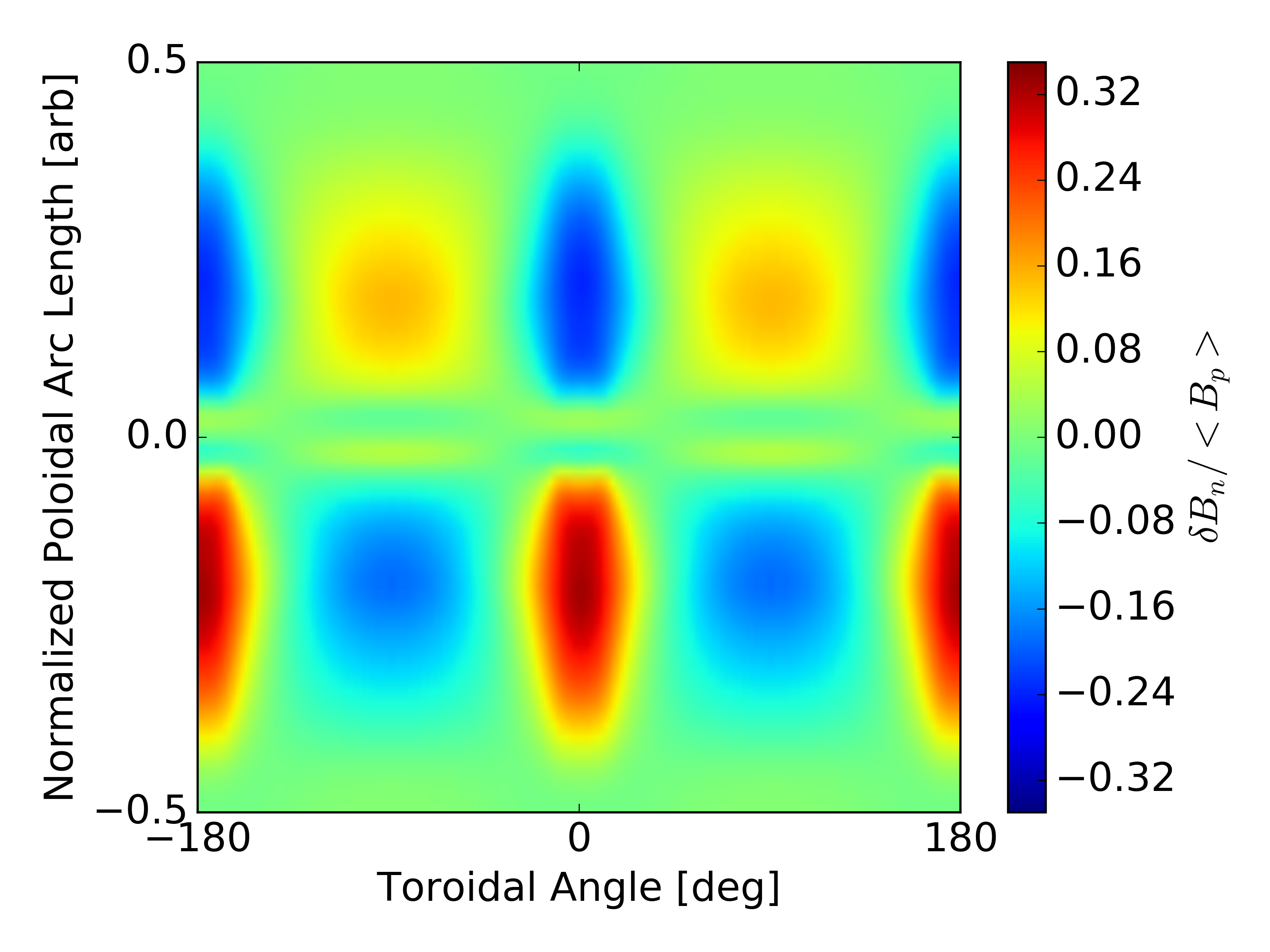}
  \end{center}
\caption{Normal field perturbation, normalized by average poloidal field, on the surface of the plasma from equilibrium
         reconstruction of LTX discharge 1504291321 at 474 ms. Zero on the y-axis corresponds to the inboard midplane,
         with positive arc length above and negative arc length below the midplane. Local magnetic field probes are located
         at 0 degrees in the toroidal coordinate. The perturbation has a predominantly
         $n=2$, $m=1$ character with the strongest perturbation associated with the toroidal shell breaks, as expected.}
\label{fig:ltx_eddy_surf}
\end{figure}

The reconstructed plasma at this time in the discharge was found to be inboard limited on the lithium-coated shell,
shown in figure \ref{fig:ltx_psi}. Strong eddy currents (colored shading in figure \ref{fig:ltx_psi}) are indicated by
the reconstruction in the VV region, with weaker eddy currents in the shell. Vertical fields produced by the
axisymmetric VV and shell eddy currents are in same directions, with both reducing the vacuum vertical
field in the plasma volume. The resulting normal field perturbation ($\delta B_n / \left< B_p\right>$) due to the
3D part of the eddy current fields
on the plasma surface is shown in figure \ref{fig:ltx_eddy_surf}. The relatively large perturbation, peaking
at $\sim$~30\% of the average poloidal field ($\sim$~5\% of the toroidal field), is primarily localized to
the toroidal shell breaks with a corresponding non-resonant $n=2$, $m=1$ character. Due to the localization to
the shell breaks, a broad mode spectrum in both poloidal and toroidal directions exists, with only even
harmonics in the toroidal direction due to symmetry of the shell.

\section{Summary and Future Work} \label{sec:discussion}
\makeatletter{}By using eddy currents with 3D vacuum fields to supplement an axisymmetric plasma equilibrium, good agreement
was achieved for reconstructions of plasma equilibria in the Lithium Tokamak eXperiment. Effects of
vacuum field screening and 3D field pickup on local magnetic diagnostics are accounted for and found
to be required for accurate reconstructions. The presented reduced model for eddy currents flowing in
the vacuum vessel and first wall shell provides a good approximation of the equilibrium impacts of
experimentally present eddy currents using only four current distributions, two for the VV region and
one for each shell set (upper/lower). This method improves on the existing reconstruction techniques
used for LTX by providing agreement with additional diagnostic sets (Thomson scattering
and the local magnetic probe array) while retaining similar or better agreement with other diagnostics.
Reliable reconstructions are crucial for discharge development and the study of confinement and transport
with a lithium first-wall --- the primary focus of the LTX program. As part of an upgrade currently underway
(LTX-$\beta$)\cite{Majeski2017}, additional diagnostics are planned to better capture 3D fields and the eddy currents
that generate the primary assymetries. In particular, measurements of the toroidal variation
of magnetic field in the shell breaks (providing direct measurement of the variation in figure
\ref{fig:ltx_eddy_surf}) are planned to better diagnose shell eddy currents.

An example equilibrium produced using this method shows fairly large 3D normal field perturbations on the
plasma surface due to 3D eddy currents in the first wall shell. The equilibrium model presented here
assumes that these perturbations are relatively small so that the total magnetic field can be represented
as an axisymmetric plasma component and a 3D vacuum component. However, given the amplitude of the normal
field perturbation, a 3D plasma state is likely. Investigation of 3D equilibria in LTX is limited by
poor toroidal distribution of presently available diagnostics; upgrades to these diagnostics are planned.
Previous attempts were also complicated by the lack of a good
axisymmetric starting point with consistent shell eddy currents\cite{Schmitt2014}. The results presented
in this paper provide both improved axisymmetric reconstructions as well as an improved starting point for
3D equilibrium reconstructions.

Further eddy current models are also being investigated based on more complete geometric models such
as VALEN\cite{Bialek2001} and CBSHL\cite{Berzak2010}. There are several effects not currently considered in the presented eddy current model,
which may be improved on by more complete models. First, the chosen eddy mode shapes do not work equally
well at all times in the discharge. This is due to the varying inductive eddy current drive throughout the
shot as coil ramp rates change and the plasma moves. As a result the true eddy current distribution, not
just the amplitude, in the shell and VV will be changing in time throughout the shot. However, most of
these sources are known resulting in known induced current distributions, so it may be possible to select
a set of eddy current modes at a given time based on known coil behaviors. As most of the coils induce
vertical field in the vacuum vessel these shapes will be similar to the chosen mode for this investigation,
so this will only be a minor correction to the presented method that already performs well. Second, 3D fields do
not come from the shell alone. The VV also has toroidal asymmetries, produced by ports, which generate
3D fields. Additionally, the VV and shell do not exist in isolation from each other but are coupled so
3D mirror currents will be induced in the VV by shell currents. The coupled method mentioned in section
\ref{sec:model} provides a method for integrating these models with equilibrium reconstruction. This method
will be tested by coupling PSI-Tri or another equilibrium code to a 3D thin-wall model.

\begin{acknowledgments}
The authors would like to thank Dr. James Bialek for many helpful discussions on eddy current modeling. Work supported
by the U.S. Department of Energy Office of Science under Contract Nos. DE-SC0016256 and DE-AC02-09CH11466.
\end{acknowledgments}

\bibliography{ltx_gs_paper}{}
\end{document}